\documentclass[%
reprint,                  
amsmath,amssymb,
aps,
]{revtex4-2}

\usepackage{graphicx}
\usepackage{dcolumn}
\usepackage{bm}

\usepackage{times}
\usepackage{hyperref}

\usepackage{comment}
\begin{document}
	
	\preprint{APS/123-QED}
	
	\title{ Thermalization Dynamics of Entanglement and non-Locality of Filtered Two-Mode Squeezed States		
	}
	
	\author{Souvik Agasti}
	\email{souvik.agasti@uhasselt.be}
	
	\affiliation{
		IMOMEC division, IMEC, Wetenschapspark 1, B-3590 Diepenbeek, Belgium
	}%
	\affiliation{
		Institute for Materials Research (IMO), Hasselt University,	Wetenschapspark 1, B-3590 Diepenbeek, Belgium
	}%

	
	\begin{abstract}

		We explore how entanglement and non-locality evolve between specific spectral components of two-mode squeezed states in thermal environments. 
		These spectral components are extracted from output modes using filters that are frequently utilized in optomechanical systems. We consider two distinct thermalization scenarios: one occurring in the vacuum state prior to entering the nonlinear crystal for squeezing, and another after the generation of the two-mode squeezed vacuum but before passing through filters and detectors. 
		Entanglement and non-locality generally remain at their peak when identical filters are applied throughout
		. In the first scenario, higher initial squeezing levels cause the dissipation of entanglement to begin slower, then accelerate over time, while the dissipation rate of non-locality moreover stays consistent. In the second scenario, greater squeezing results in a more rapid loss of both entanglement and non-locality. We identify the evolution of specific boundaries for entanglement and non-locality and the conditions for their optimization. Finally, for all the cases, increasing the thermal population of the environment enhances the rate of dissipation, whereas stronger interaction slows dissipation in a normalized dimensionless time scale.
		
	\end{abstract}
	
	\maketitle
	

	\section{Introduction}\label{Introduction}
	
	Entanglement is a fundamental aspect of quantum mechanics that plays a key role in the establishment of the next generation technological advancements such as quantum computation \cite{quantum_computation} and teleportation \cite{quantum_teleportation}, quantum information processing \cite{entanglement_quantum_cryptography} and quantum meteorology which includes gravitational wave (GW) detection \cite{study4roadmap}. The entanglement represents a bipartite correlation between two systems. A popular method to generate a continuous-variable entangled state between signal and idler beams through spontaneous parametric down-conversion process (PDC) in non-linear crystals \cite{SPDC_entanglement_generation},  which in turn produces two-mode squeezed vacuum (TMSV) states. If the systems interact with their surrounding environments, it becomes necessary to know the influence of the environment on the dynamic behavior.
	
	The dissipation dynamics of TMSV and its impact on entanglement due to its coupling to thermal reservoirs have already been studied before, both for amplitude and phase damping \cite{Decoherence_TMSV_Tohya}. In fact, thermal decoherence before \cite{Entanglement_squeezed_thermal} and after PDC \cite{Entanglement_squeezed_thermal,  Two_mode_squeezed_vacuum_common_thermal_reservoir}, both the cases have already been discussed before, investigating the conditions and limitations of entanglement. Along with that, \cite{Two_mode_squeezed_vacuum_common_thermal_reservoir} shows the thermalization dynamics of TMSV coupled to a common thermal reservoir. However, the impact of filters on their dynamics 
	has not been studied so far. Even though, we investigated the steady states of the filtered output of two-mode squeezed (TMS) thermally decohered field before in \cite{mypaper_TMSV_filter}, the dynamical behavior of entanglement has not been studied before, which therefore, becomes important to us to be focused on.

	The measurement of an observable ensures the state to be changed. In the case of a TMS state, if it is performed on one of the two parties, it also impacts the state of the other party due to entanglement. In general, local realism on a hybrid system becomes a profound feature of quantum measurement, which played a key role in solving the Einstine-Podolsky-Rosen (EPR) measurement paradox \cite{EPR_original, EPR_bell}.  The test of quantum non-locality on two-party entangled continuous-variable states remains highly important and has always been of interest in the field of quantum information science, which is still a gray area to explore. One way to test non-locality is determining the violation of Bell's inequality which is considered to be an essential condition that accounts for all local hidden variables \cite{TMSV_nonlocal_Bell_Wigner, bell_homodyne_1, bell_homodyne_2}. In this context, a few types of Bell's inequalities have been explored so far while performing homodyne measurements to test quantum nonlocality \cite{bell_homodyne_1, bell_homodyne_2}. The non-locality in the case of TMS states is examined through violation of Bell's inequality \cite{Banaszek_TMSV_bell, Banaszek_TMSV_bell_PRL}. Such type of inequality was first discussed by  Clauser, Horne, Shimony, and Holt, and therefore called CHSH inequality \cite{bell_CHSH}, which was further expressed in phase space using quasiprobability functions \cite{bell_CH}. In all cases, the two-body correlation function appears as the key element, determined from the two specially separated distant measurements.

	In this spirit, we investigate the test of the CHSH inequality on filtered TMS states.  The time evolution of nonlocality of a TMS state interacting with a thermal environment has already been well investigated before in \cite{TMSV_nonlocal_Bell_Wigner}. The thermal decoherence ensures to loss of non-local behavior of TMS states. However, the impact of the filter, applied on their output modes, on the dissipation dynamics of non-locality has never been tested so far, even though its steady-state behavior was investigated before \cite{mypaper_TMSV_filter}. Therefore, in this article, we study how filters impact on TMS state while losing its non-locality under thermal decoherence.

	Optical filters are used to select a preferable mode that has a particular frequency and a range of bandwidth. Here, we choose two different types e.g. step and exponential filters. The step filter has been in use before to determine the quantum correlation between output modes of optomechanical systems \cite{Vitali_Zippilli_NJP, entanglement_Vitali_optomechanics}. The exponential	filters were applied for the time-dependent evolution of the spectrum of light \cite{filter_2_JOSA} and also in feedback-controlled nanoscale optomechanical systems \cite{Asjad_Vitali_filter_2}.
	
	In this article, we study the dynamics of entanglement and non-locality of the filtered output of thermally decohereted TMS states. We consider two cases of thermalization. The first one is wideband vacuum lights are being thermalized before entering into the non-linear crystal for the PDC process, i.e. two-mode squeezed thermally decohered field (TMSTDF). The second one is the thermal decoherence occurring after the PDC process, which therefore generates a thermally decohereted two-mode squeezed vacuum (TDTMSV). In the following section, we establish the theory of thermalization dynamics of TMSTDF and study its impact on entanglement and non-locality. Afterward, we follow similar studies for TDTMSV, with a comparative discussion.
	

	\begin{figure}[t!]
		\includegraphics[width= 1 \linewidth]{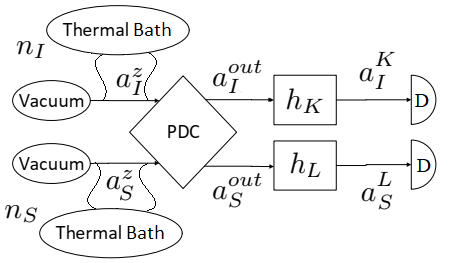}
		\caption{ Block diagram of the detection of filtered TMSTDF. The thermally decohered input vacuum goes through the parametric down-conversion (PDC) process. Optical filters are applied on two-mode squeezed output before being detected at D.  }\label{Block_Diagram_thermal}
	\end{figure}


	\begin{figure}
		\includegraphics[width= 1 \linewidth]{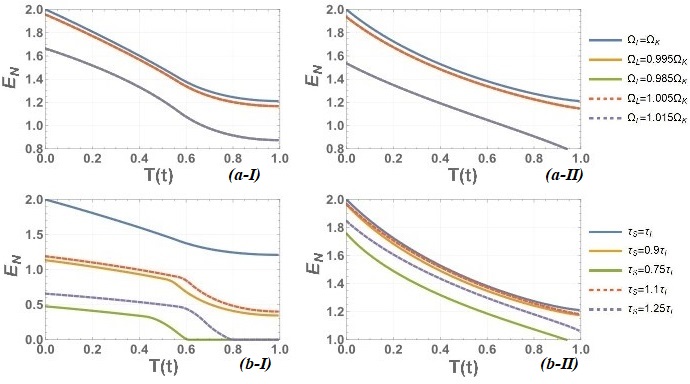}
		\caption{ Entanglement between filtered outputs for the variation of
			parameters of (a) central frequencies for $\tau_I = \tau_S = 0.2\Omega_K$ (b)
			filter linewidths for $\Omega_L = \Omega_K $ and $\tau_I = 0.2\Omega_K$, for (I) step filter and (II) exponential filters. Fixed parameters are $r = 1, n_1 = n_2 = 0.6, \kappa^{in}_I= \kappa^{in}_S = \kappa= 0.07 \Omega_K.$
		}\label{Ent_W_t}
	\end{figure}

	\begin{figure*}
		\includegraphics[width= 1 \linewidth]{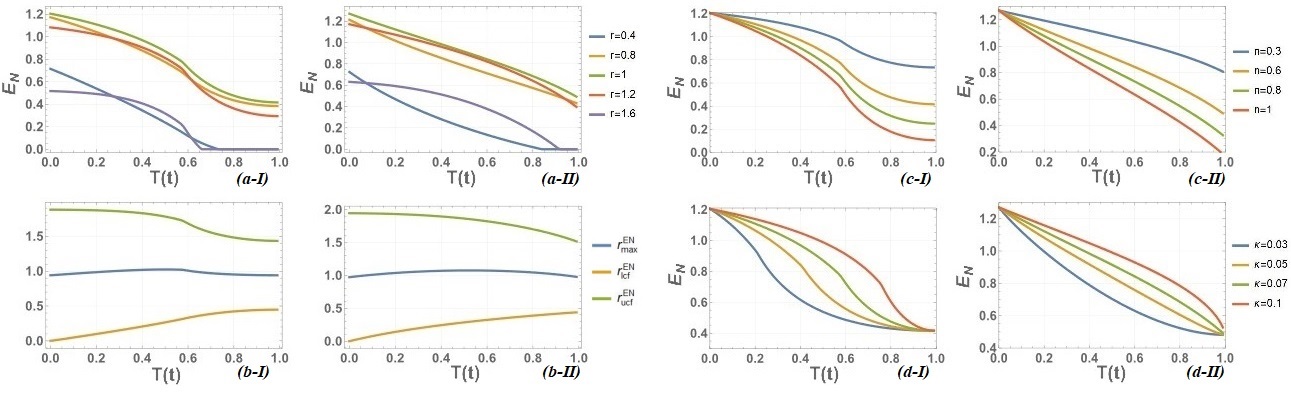}
		\caption{ Entanglement between filtered outputs for the variation of (a) $r$, (c) $n_1=n_2 = n$ and (d) $\kappa^{in}_I =\kappa^{in}_S = \kappa$ for (I) step filter and (II) exponential filters. In (b): maximal entanglement and its cutoff limits. Fixed parameters are $\Omega_L = 1.02\Omega_K, \tau_I = 0.2\Omega_K, \tau_S = 0.208 \Omega_K, r = 1, n_1 = n_2 = 0.6, \kappa^{in}_I= \kappa^{in}_S = 0.07 \Omega_K.$.
		}\label{Ent_R_n_k}
	\end{figure*}

	\begin{figure}
		\includegraphics[width= 1 \linewidth]{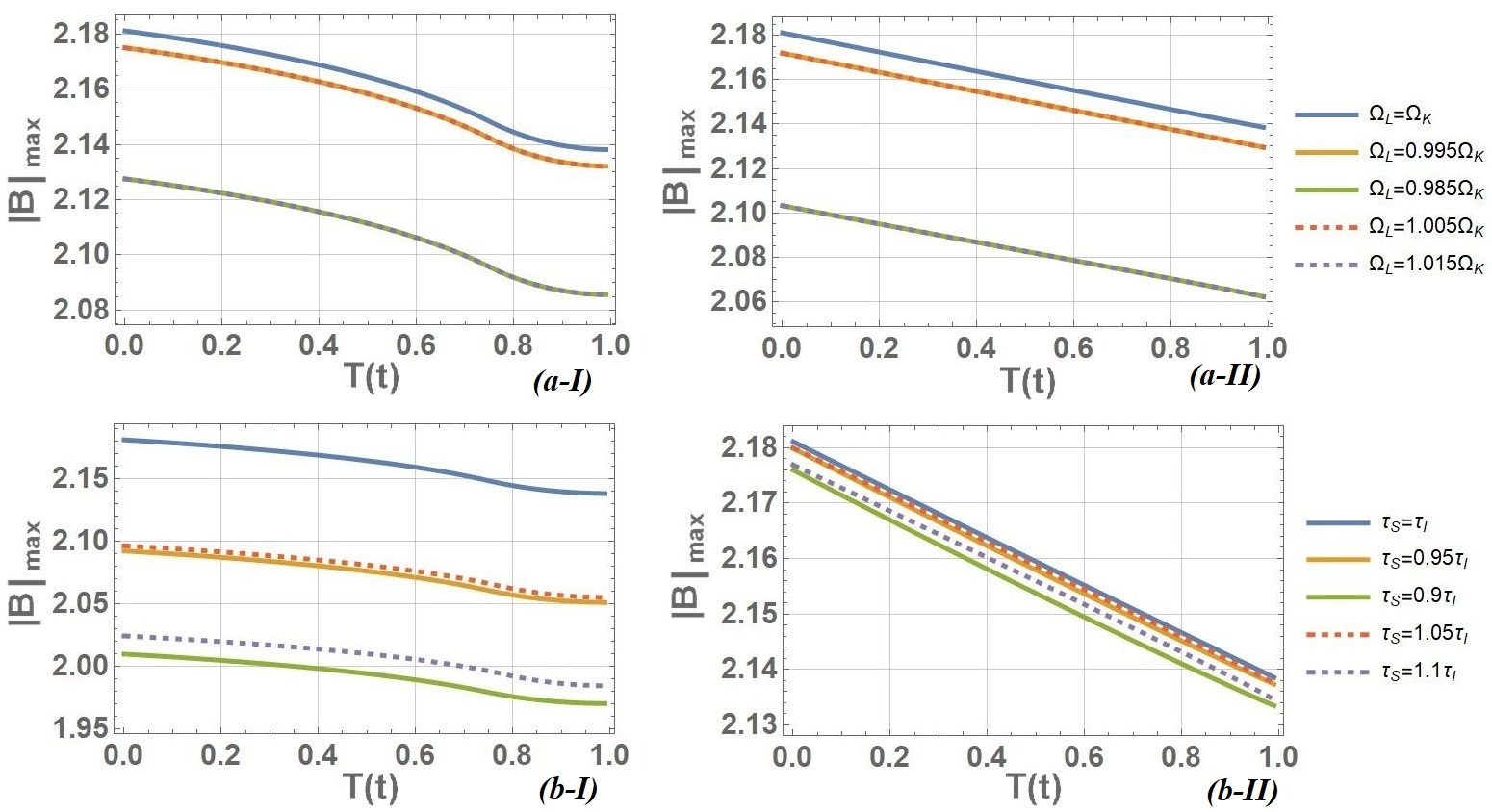}
		\caption{ The maximal value of the Bell function for the variation of (a) central frequencies for $\tau_I = \tau_S = 0.2\Omega_K$ (b) filter linewidths for $\Omega_L = \Omega_K $ and $\tau_I = 0.2\Omega_K$, for (I) step filter and (II) exponential filters. Fixed parameters are $
			r= 0.4, n_1 = n_2 = 0.1, \kappa^{in}_I= \kappa^{in}_S = 0.1 \Omega_K.$ 
		}\label{Bell_W_t}
	\end{figure}

	\begin{figure*}
		\includegraphics[width= 1 \linewidth]{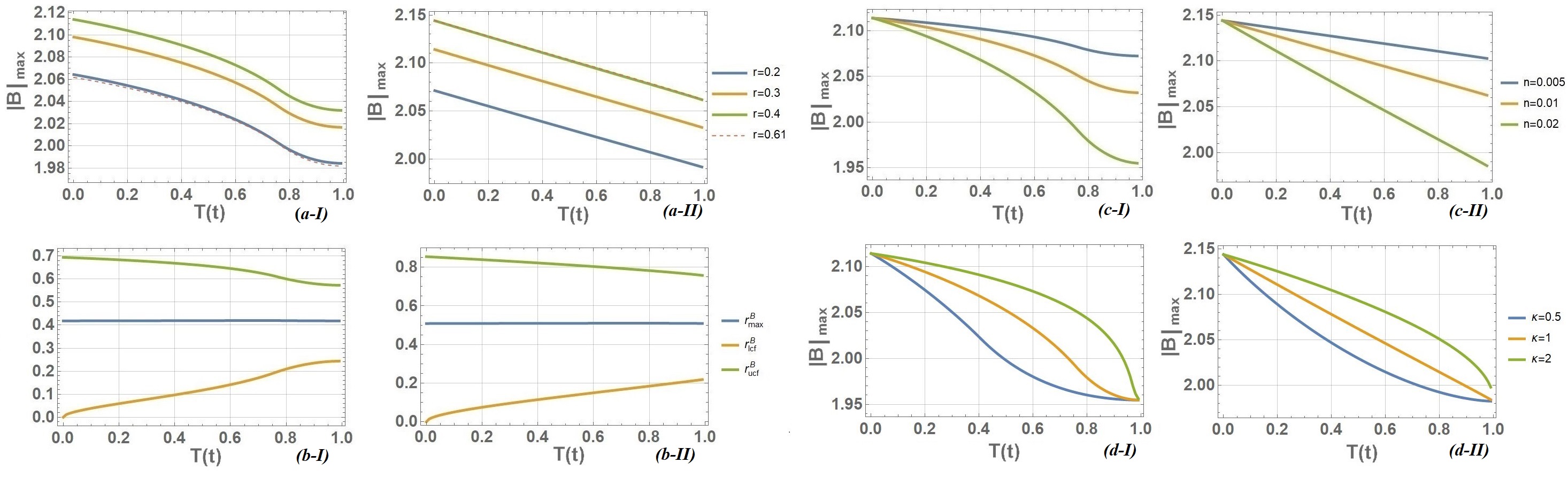}
		\caption{ The maximal value of Bell function for the variation of (a) $r$, (c) $n_1=n_2 = n$ and (d) $\kappa^{in}_I =\kappa^{in}_S = \kappa$ for (I) step filter and (II) exponential filters. In (b): maximal non-locality and its cutoff limits. Fixed parameters are $\Omega_L = 1.01\Omega_K, \tau_I = 0.2\Omega_K, \tau_S = 0.205 \Omega_K, r= 0.4,  n_1 = n_2 = 0.1, \kappa_I= \kappa_S = 0.1 \Omega_K$ . 
		}\label{Bell_R_n_k}
	\end{figure*}

	\section{FILTER on TMSTDF}

	\subsection{Input TMS states}

	Entangled TMS states are produced through PDC process in non-linear optical crystals $(z)$ shown in Fig. \ref{Block_Diagram_thermal}. The figure also shows the interaction of the input vacuum to its corresponding reservoirs before entering into the non-linear crystal. The squeezing through the PDC process can be represented by the  Bogliubov modes as

	\begin{align}
		\begin{bmatrix}
			X_I^{out}\\
			Y_I^{out}\\
			X_S^{out}\\
			Y_S^{out}
		\end{bmatrix} &= \begin{bmatrix}
			\cosh r & 0 & \sinh r & 0 \\
			0 & \cosh r & 0 & - \sinh r \\
			\sinh r & 0 & \cosh r & 0\\
			0 & - \sinh r & 0 & \cosh r
		\end{bmatrix} \begin{bmatrix}
			X_I^{z}\\
			Y_I^{z}\\
			X_S^{z}\\
			Y_S^{z}
		\end{bmatrix} 
	\end{align}
	
	where $r$ is the amplitude of the squeezing parameter and the arbitrary phase of squeezing is fixed to $\pi/2$.	
	$X_{I,S}^{z}(t) = \frac{1}{\sqrt{2}} \left(a_{I,S}^{z}(t) +{a_{I,S}^{z} }^\dagger(t) \right) $ and $Y_{I,S}^{z}(t) = -i\frac{1}{\sqrt{2}} \left(a_{I,S}^{z}(t) -{a_{I,S}^{z} (t)}^\dagger \right) $ represents the amplitude and phase quadratures of wideband bosonic modes, respectively, which are directed to the non-linear crystal (z) as inputs of the idler $(I)$ and signal $(S)$ modes. $a_{I,S}^{z}$ and ${a_{I,S}^{z} }^\dagger$ represent corresponding annihilation and creation operators.  
	The outputs of the non-linear crystal generate a TMS state of amplitude and phase quadratures $X_{I,S}^{out}$ and $Y_{I,S}^{out}$. Likewise vacuum modes, corresponding annihilation and creation operators are $a_{I,S}^{out} $ and ${a_{I,S}^{out} }^\dagger $, respectively, for the outputs of systems $I$ and $S$; which gives	 
	$X_{I,S}^{out}(t) = \frac{1}{\sqrt{2}} \left(a_{I,S}^{out}(t) +{a_{I,S}^{out} }^\dagger(t) \right) $ and $Y_{I,S}^{out}(t) = -i\frac{1}{\sqrt{2}} \left(a_{I,S}^{out}(t) -{a_{I,S}^{out} (t)}^\dagger \right) $.	  
	The Hamiltonian that describes the interaction of input vacuum modes to their corresponding thermal reservoir is
	
	\begin{equation}
		H^z_{I,S} = \int_{-\infty}^{\infty} \eta^z_{I,S}(\omega) \left(a^z_{I,S} b^z_{I,S} (\omega)+h.c. \right) \mathrm{d}\omega
	\end{equation} 
	
	where $b^z_{I,S} (\omega)$ represents the mode of the corresponding reservior, and $\eta^z_{I,S}(\omega)$ is the coupling strength for the corresponding mode. Under wide band limit approximation $\eta^z_{I,S}(\omega) = \sqrt{\kappa^{in}_{I,S}/\pi}$, which determines rate of thermalization.

	\subsection{Filtered Output Modes}
	
	Since the field is continuous, we can extract independent optical modes for  different time intervals, as expressed in Fig. \ref{Block_Diagram_thermal}:
	
	\begin{equation} \label{filter_out_mode}
		a^{K,L}_{I,S} (t) 
		=  \int_{-\infty }^t \mathrm{d}t'	h_{K,L}(t-t') a^{out}_{I,S} (t')
	\end{equation}
	
	where
	$h_{K,L}(t) $ are the filter functions of the $K,L$th output modes of the parties I and S, respectively. Before applying the filter, the regular output field obeys the correlation function: $[a_{k}^{out}(t),{a_{k}^{out} }^\dagger(t')] = \delta(t-t') $ where $k\in \{I,S\}$. 
	Equivalently, the commutation relation continues to be followed for a generic set of output modes, even after applying a filter: $[a_{k}^{M} ,{a_{k}^{N}}^\dagger] = \delta_{MN}$. This is  ensured only when the orthonormality between modes is followed, which is
	
	\begin{equation}
		\int_{0}^\infty  {h}_i(t) {h}^*_j(t) \, \mathrm{d}t = \delta_{ij}
	\end{equation}
	
	The filtered output modes, given in Eq. \eqref{filter_out_mode}, can be transformed into the frequency domain, which also satisfies orthogonality relations. 

	Here, we pick up explicit sets of two different types of filter functions that follow such orthogonality. The type-I is a step filter function, given by
	
	\begin{equation}\label{step_filter_t}
		h_{K,L}(t) = \frac{ \Theta(t) - \Theta(t-\tau_{I,S}) }{\sqrt{\tau_{I,S} }} e^{-i \Omega_{K,L} t }
	\end{equation}
	
	$\Theta(t)$ is the Hevisite step function. The filter function offers a set of independent optical modes, which are located around the central frequencies $(\Omega_{K,L})$. 
	$\tau_{I,S}$ defines time interval of corresponding systems $I$ and $S$, which therefore gives the spectral width $\tau_{I,S}^{-1}$, which defines seperation between mode frequencies as
	
	\begin{equation}\label{periodicity}
		\Omega_{K}-\Omega_{K \pm n} = \pm n\frac{2 \pi}{\tau_{I} }  \hspace{1 mm} \text{and} \hspace{1 mm} \Omega_{L}-\Omega_{L \pm n} = \pm n\frac{2 \pi}{\tau_{S} },  \hspace{1 mm}  n \, \text{integer}
	\end{equation}

	Such functions were used before in \cite{entanglement_Vitali_optomechanics, Vitali_Zippilli_NJP} to filter the output modes of optomechanical systems. 

	We also consider another type of filter (type-II) which is based upon an exponential function:
	
	\begin{equation}\label{exponential_filter_t}
		{h}_{K,L}(t) = \frac{e^{-(1/\tau_{I,S} + i\Omega_{K,L})t}}{\sqrt{\tau_{I,S}/2}} \Theta(t) ,
	\end{equation}
	which has the same periodicity as Eq. \eqref{periodicity}. Exponential filters have been used before in optomechanical systems \cite{Asjad_Vitali_filter_2, filter_2_JOSA} for the analysis of the time-dependent spectrum of light.

	\subsection{Time-dependent correlation matrix}

	We first determine the generalized expression for the filtered output, which further determines the correlation matrix. For both the step and exponential filters, one can obtain an infinite number of mutually independent filtered output quadratures, by selecting their frequencies and bandwidths. The filtered TMS quadratures are
	
	\begin{align} \label{quadratures_sqz_freq}
		\begin{bmatrix}
			X_{I,S}^{K,L} (r; t) \\
			Y_{I,S}^{K,L} (-r;t) 
		\end{bmatrix} &= \int_{-\infty}^{t} \mathrm{d}t' {T}_{K,L} (t-t') \begin{bmatrix}
			X_{I,S}^{out} (r;t') \\
			Y_{I,S}^{out} (-r;t')
		\end{bmatrix} 
	\end{align}

	where
	
	\begin{equation} \label{T_mat_freq}
		{T}_{K,L}(t) =\begin{bmatrix}
			\Re({h}_{K,L}) & -\Im({h}_{K,L})\\
			\Im({h}_{K,L}) & \Re({h}_{K,L})
		\end{bmatrix} 
	\end{equation}

	and  $X^{K,L}_{I,S}(r;t) = (a_{I,S}^{K,L}(r;t)+ {a_{I,S}^{K,L}(r;t) }^\dagger)/\sqrt{2}, Y^{K,L}_{I,S}(r;t) = -i (a_{I,S}^{K,L}(r;t)- {a_{I,S}^{K,L} }^\dagger(r;t)  )/\sqrt{2}$ represents dimensionless amplitude and phase filtered quadrature operators. The elements of the correlation matrix of $V(r)$ are defined by
	
	\begin{equation}\label{vmat_eliments}
		{V_{ij} (r)} = \frac{1}{2} \langle v_{i}^{out} v_{j}^{out} + v_{j}^{out} v_{i}^{out} \rangle
	\end{equation}
	
	where

	\begin{equation}
		v^{out} = [X_I^{K}(r), Y_I^{K}(r), X_S^{L}(r), Y_S^{L}(r) ]^T
	\end{equation}
	
	Selecting the elements for the squeezing factor $+r$ from \eqref{quadratures_sqz_freq}, we build up the correlation matrix  as $
	V (r;t) =  \begin{bmatrix}
		V_I & V_{corr}^T\\
		V_{corr} & V_S 
	\end{bmatrix}$, 
	where
	
	\begin{subequations} \label{correlation_matrix}
		\begin{align}
			V_I(t) &= \frac{1}{2} \text{Diag}[D_I ,\, D_I ], \\ 
			V_S(t) &= \frac{1}{2} \text{Diag}[D_S ,\, D_S ] \, \text{and} \\
			V_{corr}(t) &= \frac{1}{2}\begin{bmatrix}
				\begin{array}{cccc}
					C_{11} & C_{12} \\
					C_{21} & C_{22} 
				\end{array}
			\end{bmatrix}, 
		\end{align}
	\end{subequations}
	
	The elements of the steady-state correlation matrix for each mode of the filtered output of the bipartite system are
	
	\begin{subequations}
		\begin{align}\label{mat_elmnt_intgrl}
			D_I &= \left(\Im({h}_K^2)+\Re({h}_K^2) \right) \star \left[ (p_I+p_S) \cosh (2 r)+p_I-p_S \right] \\
			D_S &= \left(\Im({h}_L^2)+\Re({h}_L^2)\right) \star \left[(p_I+p_S) \cosh (2 r)-p_I+p_S) \right]\\
			C_{11} &= \left(\Im({h}_K) \Im({h}_L)+ \Re({h}_K) \Re({h}_L) \right) \star \left[p_I+ p_S\right] \sinh (2 r) \\
			C_{12} &=  \left( \Im({h}_K) \Re({h}_L)- \Im({h}_L) \Re({h}_K) \right) \star \left[ p_I- p_S \right] \sinh (2r) \\
			C_{22} &= -C_{11}  ;  C_{21} =  C_{12}
		\end{align}
	\end{subequations}
	
	where
	
	\begin{subequations}
		\begin{align}
			p_I (t) &= n_I\Theta(t) \left(1-e^{-2 \kappa_I t}\right)+ \frac{1}{2} \\
			p_S (t) &= n_S\Theta(t) \left(1-e^{-2 \kappa_S t}\right)+ \frac{1}{2}
		\end{align}
	\end{subequations}
	
	and 
	$A \star B = \int_{-\infty }^{t} \mathrm{d}t' A(t-t') B(t')  $ represents convolution between two entities. $\Theta(t)$ indicates the time when the interaction with the environment is switched on.	One can check the elements and compare with \cite{Entanglement_squeezed_thermal} when no filter is applied. 
	The elements of the correlation matrix in Eq. \eqref{mat_elmnt_intgrl} can be solved to

	\begin{widetext}	
		\begin{subequations}
			\begin{align}\label{mat_elmnt}
				D_I &= n_I \left[ I_I(0)- e^{-2\kappa_It} I_I(\kappa_I)  \right] (1+\cosh (2 r)) -  n_S\left[I_I(0)-e^{-2\kappa_St} I_I(\kappa_S)  \right] (1-\cosh (2 r)) +\cosh (2 r) \\
				D_S &= -n_I \left[ I_S(0)-e^{-2\kappa_It}I_S(\kappa_I)  \right] (1-\cosh (2 r)) +  n_S\left[I_S(0)-e^{-2\kappa_St}I_S(\kappa_S)  \right] (1+\cosh (2 r)) +\cosh (2 r) \\
				C_{11} &=-C_{22} = \sinh (2 r) \left[ -n_I e^{-2\kappa_It} J_c(\kappa_I) - n_S e^{-2\kappa_St} J_c(\kappa_S ) + (n_I + n_S )J_c(0) + K_f \right]      \\
				C_{12} &=  C_{21} = - \sinh (2 r) \left[ -n_I e^{-2\kappa_It} J_s(\kappa_I) + n_S e^{-2\kappa_St} J_s(\kappa_S ) + (n_I - n_S )J_s(0) \right]   
			\end{align}
		\end{subequations}
		
		where

		\begin{subequations}
			\begin{align}
				J_s(\kappa) &=  \frac{ e^{2 \kappa  \tau} \left[2 \kappa  \sin \left(\tau \left(\Omega _K-\Omega _L\right)\right)+\left(\Omega _L-\Omega _K\right) \cos \left(\tau \left(\Omega _K-\Omega _L\right)\right)\right]+ \Omega _K-\Omega _L }{\sqrt{\tau _I \tau _S} \left(4 \kappa ^2+\left(\Omega _K-\Omega _L\right){}^2\right)}\\
				J_c(\kappa) &= \frac{ e^{2 \kappa  \tau} \left[ 2 \kappa  \cos \left(\tau \left(\Omega _K-\Omega _L\right)\right)+\left(\Omega _K-\Omega _L\right) \sin \left(\tau \left(\Omega _K-\Omega _L\right)\right)\right]-2 \kappa }{\sqrt{\tau _I \tau _S} \left(4 \kappa ^2+\left(\Omega _K-\Omega _L\right){}^2\right)} 
			\end{align}
		\end{subequations}

		for filter-I where $\tau = \min [t,\tau_{I}, \tau_{S}]$, and in case of filter-II

		\begin{subequations}
			\begin{align}
				&J_s(\kappa) = \frac{2}{{\sqrt{\tau_{I} \tau_{S}}}} \left[ \frac{e^{(2 \kappa -\frac{1}{\tau_I} - \frac{1}{\tau_S} )t} \left( (2\kappa - \frac{1}{\tau_I} - \frac{1}{\tau_S} ) \sin \left(t (\Omega _K-\Omega _L )\right)+\left(\Omega _L-\Omega _K\right) \cos \left(t \left(\Omega _K-\Omega _L\right)\right)\right) + \left(\Omega _K-\Omega _L\right)}{ (2\kappa - \frac{1}{\tau_I} - \frac{1}{\tau_S} )^2+\left(\Omega _K-\Omega _L\right){}^2} \right]\\
				&J_c(\kappa) = \frac{2}{{\sqrt{\tau_{I} \tau_{S}}}} \left[ \frac{e^{ (2 \kappa -\frac{1}{\tau_I} - \frac{1}{\tau_S} )t} \left( (2\kappa - \frac{1}{\tau_I} - \frac{1}{\tau_S} )  \cos \left(t (\Omega _K-\Omega _L )\right)+\left(\Omega _K-\Omega _L\right) \sin \left(t \left(\Omega _K-\Omega _L\right)\right)\right) - (2 \kappa -\frac{1}{\tau_I} - \frac{1}{\tau_S} ) }{(2\kappa - \frac{1}{\tau_I} - \frac{1}{\tau_S} )^2+\left(\Omega _K-\Omega _L\right){}^2} \right],
			\end{align}
		\end{subequations}
	\end{widetext}

	\begin{subequations}
		\begin{align}
			I_{I,S} (\kappa) &= \frac{ e^{2 \kappa  \tau }- 1 }{2 \kappa  \tau_{I,S}} \,\, \text{where $\tau = \min [t,\tau_{I,S}]$ for filter-I} \\
			I_{I,S} (\kappa) &= \frac{1}{ \tau_{I,S} } \left[ \frac{e^{2( \kappa -\frac{1}{\tau_{I,S}} )t} - 1 }{( \kappa - \frac{1}{\tau_{I,S}} ) } \right] \,\, \text{for filter-II,}
		\end{align}
	\end{subequations}	
	
	and
	
	\begin{subequations}	
		\begin{align}
			K_f  & = \frac{\sin [\tau(\Omega_{K} - \Omega_{L})] }{\sqrt{\tau_I \tau_S} (\Omega_{K} - \Omega_{L}) } \,\, \text{where} \, \, \tau = \min [\tau_I,\tau_S] \text{ for filter-I} \\
			K_f &= \frac{2}{{\sqrt{\tau_{I} \tau_{S}}}} \left[ \frac{ (\frac{1}{\tau_I} + \frac{1}{\tau_S} )}{ ( \frac{1}{\tau_I} + \frac{1}{\tau_S} )^2+\left(\Omega _K-\Omega _L\right){}^2} \right]   \,\, \text{for filter-II}
		\end{align}
	\end{subequations}
	
	$K_{f}$ is unit valued when the filters are identical and it drops down when they are non-identical, i.e. $\Omega_K \neq\Omega _L$ and $\tau_I \neq \tau_S$, for both the types of filters.

	\begin{figure}[t!]
		\includegraphics[width= 1 \linewidth]{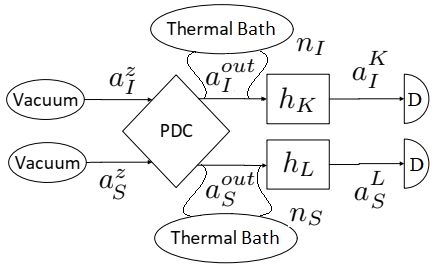}
		\caption{ Block diagram of the detection of filtered TDTMSV. TMSV is generated using the parametric down-conversion (PDC) process, and the thermalization happens afterwards. Optical filters are applied on two-mode squeezed output before being detected at D. }\label{Block_Diagram_Thermal_decohr}
	\end{figure}


	\begin{figure}
		\includegraphics[width= 1 \linewidth]{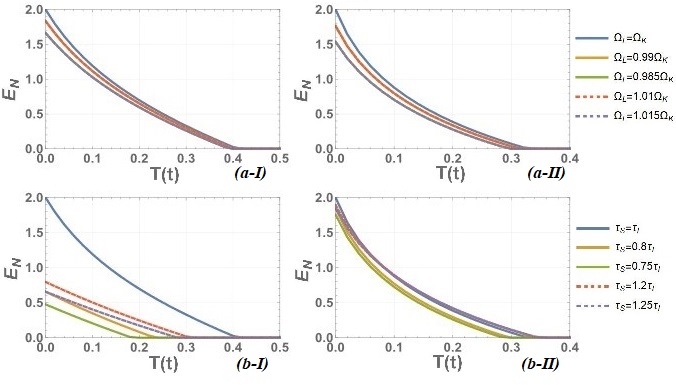}
		\caption{ Entanglement between filtered outputs for the variation of
			parameters of (a) central frequencies for $\tau_I = \tau_S = 0.2\Omega_K$ (b)
			filter linewidths for $\Omega_L = \Omega_K $ and $\tau_I = 0.2\Omega_K$, for (I) step filter and (II) exponential filters. All other fixed parameters remains same with Fig. \ref{Ent_W_t}. 
		}\label{LD_Ent_W_t}
	\end{figure}

	\begin{figure*}
		\includegraphics[width= 1 \linewidth]{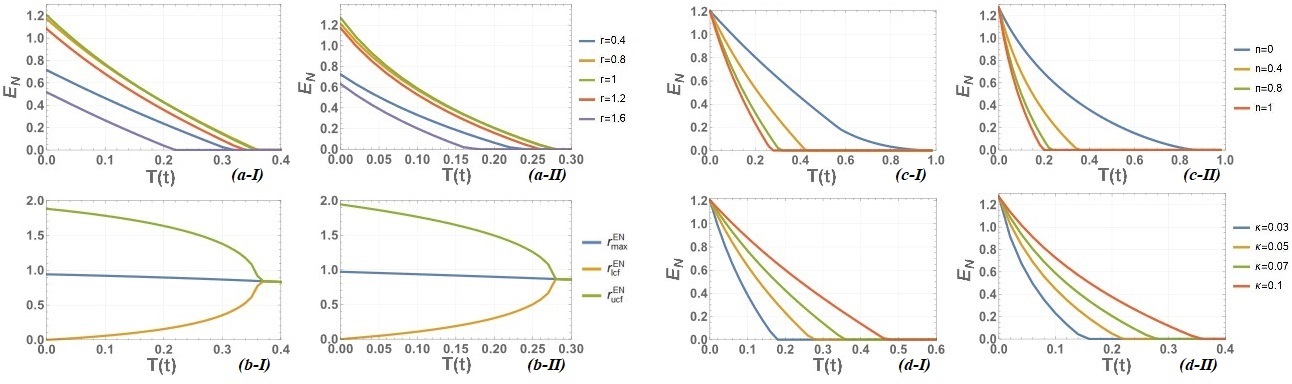}
		\caption{ Entanglement between filtered outputs for the variation of (a) $r$, (c) $n_1=n_2 = n$ and (d) $\kappa^{out}_I =\kappa^{out}_S = \kappa$ for (I)step filter and (II) exponential filters.  In (b) maximal entanglement and its cutoff limits. All other fixed parameters remain the same with Fig. \ref{Ent_R_n_k}. 
		}\label{LD_Ent_R_n_k}
	\end{figure*}

	\begin{figure}
		\includegraphics[width= 1 \linewidth]{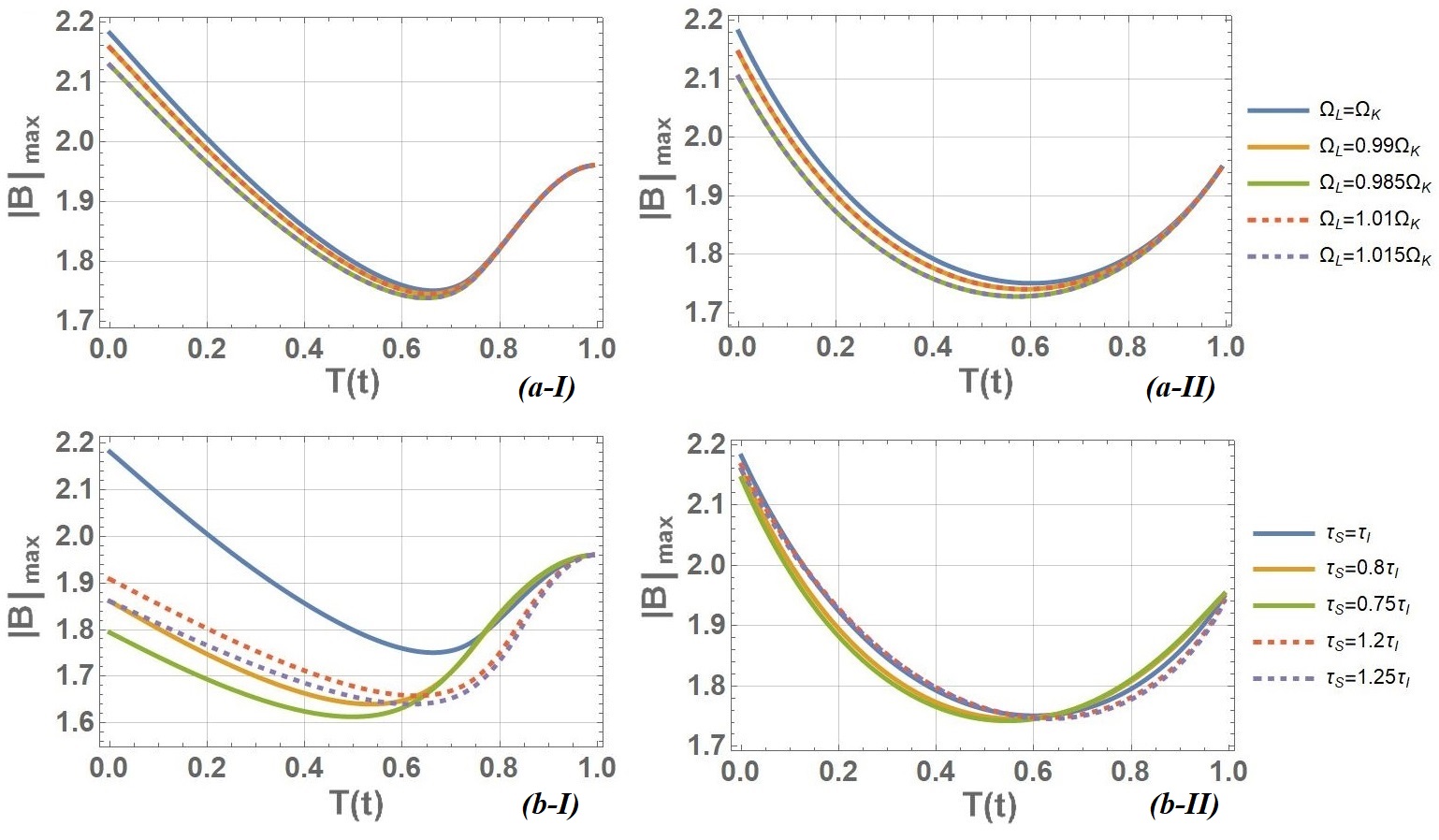}
		\caption{ The maximal value of the Bell function for the variation of (a) central frequencies for $\tau_I = \tau_S = 0.2\Omega_K$ (b) filter linewidths for $\Omega_L = \Omega_K $ and $\tau_I = 0.2\Omega_K$, for (I) step filter and (II) exponential filters. All other fixed parameters remain the same with Fig. \ref{Bell_W_t}.
		}\label{LD_Bell_W_t}
	\end{figure}

	\begin{figure*}
		\includegraphics[width= 1 \linewidth]{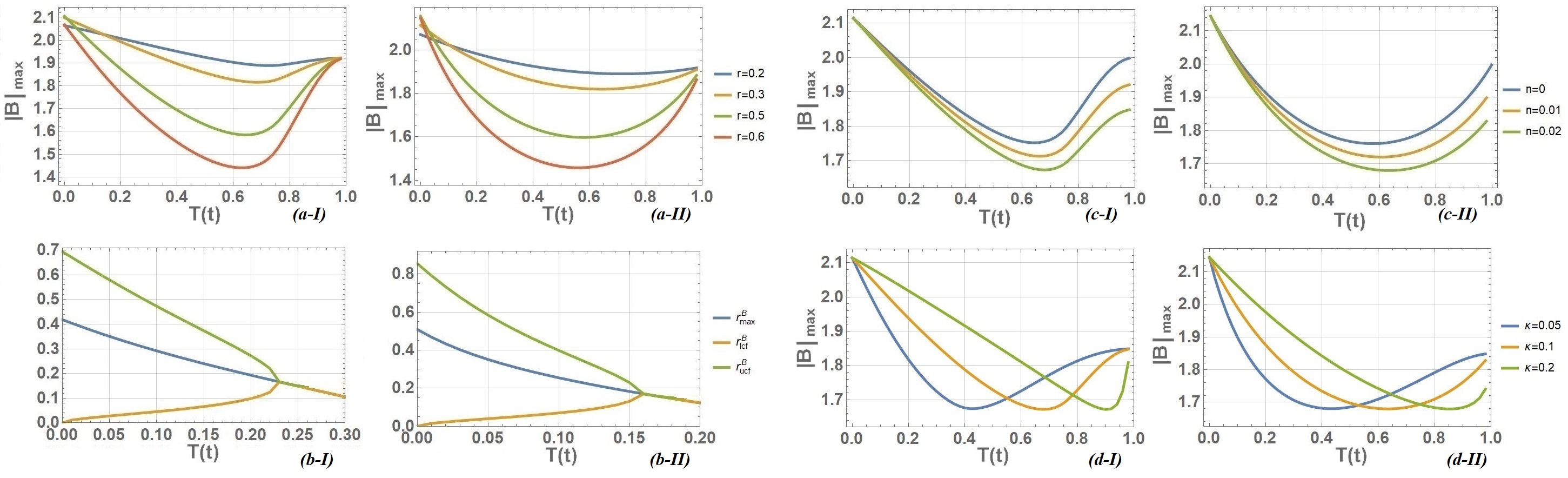}
		\caption{ The maximal value of Bell function for the variation of (a) r, (c) $n_1=n_2 = n$ and (d) $\kappa^{out}_I =\kappa^{out}_S = \kappa$ for (I)step filter and (II) exponential filters. In (b): maximal non-locality and its cutoff limits. All other fixed parameters remain the same with Fig. \ref{Bell_R_n_k}.
		}\label{LD_Bell_R_n_k}
	\end{figure*}

	\subsection{Two-Mode Entanglement- Logarithmic Negativity}
	
	The entanglement between two parties can be witnessed by determining the quantity \cite{Simon}
	
	\begin{equation}\label{entanglement_description}
		E_N = \max[0, -\ln 2\nu^-],
	\end{equation}
	
	where
	
	\begin{equation}
		\nu^- = \sqrt{\frac{ \Sigma (V) + \sqrt{\Sigma (V)^2 - 4 \det (V)} }{2}}
	\end{equation}
	
	where $ \Sigma(V ) = ( \det V_I + \det V_S - 2 \det V_{corr} )$.

	The entanglement is ultimately influenced by the system and filter parameters. In Fig. \ref{Ent_W_t}, we illustrate the dissipation dynamics of entanglement plotted in a normalized time scale $(T(t)= 1-\exp(-\kappa t))$,  and discuss how it varies with different filter settings. As discussed earlier in \cite{mypaper_TMSV_filter} the steady-state entanglement reaches its maximum when the filters are identical $(\Omega_{K} = \Omega_{L}$ and $\tau_{I} = \tau_{S})$; the thermal decoherence of entanglement does not show any exemption here. Following the steady-state pattern, entanglement uniformly decreases when there is a mismatch in the central frequencies $(\Omega_{K} \neq \Omega_{L})$, as shown in Fig. \ref{Ent_W_t}(a), regardless of the filter type. However, in contrast to central frequencies, Fig. \ref{Ent_W_t}(b) demonstrates that when the bandwidths are mismatched $(\tau_{I} \neq \tau_{S})$, the reduction in entanglement does not remain uniform.

	The most significant feature of non-uniformly filtered two-mode squeezed (TMS) states is the decrease in entanglement with further increases in the input degree of squeezing $(r)$, following an initial rise, which forms a bell-shaped pattern \cite{mypaper_TMSV_filter}. Interestingly, this reduction in entanglement with increasing $r$ may not persist throughout the entire thermalization period of the input vacuum. Exceptions to this are seen in Fig. \ref{Ent_R_n_k}(a), where both types of filters demonstrate that entanglement can become more resilient at higher values of $r$ during thermal decoherence, even though the initial and steady-state entanglement is weaker. The dissipation of entanglement slows down at the beginning for higher $r$ and accelerates over time, for both the types of filters. 
	This occurs because the filtering interval covers a much longer period than the period of thermalization towards the beginning. This phenomenon leads to another event which is explained in Fig. \ref{Ent_R_n_k}(b), which shows that the initial degree of squeezing at which entanglement reaches its maximum $(r^{EN}_{max})$ rises initially during the early stages of time evolution and declines further. Additionally, we observe that the lower and upper cutoff limits of the squeezing parameters $(E_N (r^{EN}_{lcf}\le r \le r^{EN}_{ucf}) > 0)$ are narrowing the range where entanglement is sustained.

	The dissipation of entanglement is enhanced with the increment of the thermal population of the reservoir and the coupling strength.
	Fig. \ref{Ent_R_n_k}(c) ensures the phenomenon for both types of filters, anticipating the steady-state entanglement to reduce for higher thermal population \cite{mypaper_TMSV_filter}. Note that, even though increasing the coupling constant increases the rate of dissipation, the entanglement decays slower w.r.t. the dimensionless rescaled time: $T(t) = 1-\exp(-\kappa t)$  (Fig. \ref{Ent_R_n_k}(d)). This happens due to the fact that the coupling constant changes the time scale, which eventually rescales the central frequency and the linewidths of the filter. Reduction of coupling constant effectively makes the filters more non-identical, i.e. it rescales their effective linewidths which increases discrepancy between them.

	\subsection{Quantum Non-Locality}
	
	The non-locality of the filtered TMS state even though has been studied before in \cite{mypaper_TMSV_filter}, the thermalization dynamics has not been discussed yet. In the cases of pure states, the entanglement defines nonlocality. However, as the thermalization makes the state mixed, the entanglement therefore becomes necessary, but not sufficient for non-locality. The violation of Bell's inequality which justifies non-locality, is measured from the maximum value of the Bell's function $(|B|_{max})$. One can express it in terms of Wigner functions as	\cite{Banaszek_TMSV_bell, TMSV_nonlocal_Bell_Wigner}
	
	\begin{equation}\label{bell_func_deff}
		B = \frac{\pi^2}{4} \left[ W(u_M^{00}) +W(u_M^{01}) W(u_M^{10}) - W(u_M^{11})          \right]
	\end{equation}
	
	where 
	
	\begin{equation}
		W( {u^{mn}_M}) = \frac{1}{ \pi^2 \sqrt{\det[ {V(r)}]}} \exp \left[ -\frac{1}{2} {u^{mn}_M}^T {V(r)}^{-1} {u^{mn}_M}\right]
	\end{equation}
	
	represents the Wigner function for  ${u^{mn}_M} = [Q^m_I, P^m_I, Q^n_S, P^n_S ]^T $ is a vector in phase space. 
	The quantum mechanical description of a field to be nonlocal when condition $|B|_{max}\leq 2$ violates. Larger $|B|_{max}$ indicates non-locality to be stronger. 
	Here, we provide a numerical estimation of $|B|_{max}$ which shows the impact of non-identical filters on the thermalization dynamics of the TMSTDF.

	The Bell function decays down throughout the thermalization process of the TMS field, which eventually did not change even after applying the filter. 
	Likewise, entanglement, the reduction of Bell function occurs uniformly with the mismatch of filter frequencies (Fig. \ref{Bell_W_t}(a)) and non-uniformly for the mismatch of filter linewidths, throughout the period of thermalization and irrespective of the type of filters. We noticed before in \cite{mypaper_TMSV_filter} that the Bell function increases initially and further drops down with the initial degree of squeezing $(r)$. Fig. \ref{Bell_R_n_k}(a) shows that the phenomenon remains unchanged throughout the evolution. Therefore, unlike entanglement, the input degree of squeezing for which the Bell function becomes maximum ($r^B_{max})$ remains moreover steady throughout the evolution (Fig. \ref{Bell_R_n_k}(b)). As anticipated, the lower and upper cutoff limits $(|B|_{max}(r_{lcf} \le r \le r_{ucf}) > 2)$ shows the region of nonlocality squeezes during evolution. Besides, likewise entanglement, the Bell function also dissipates faster for higher thermal population (Fig. \ref{Bell_R_n_k}(c)) and slower for stronger coupling with the environment (Fig. \ref{Bell_R_n_k}(d)) w.r.t. dimensionless rescaled time, for both the filters.

	Moreover, the non-locality appears to be within the region of entanglement, which has been justified before from the mixedness of states	\cite{mypaper_TMSV_filter, mixed_entanglement_nolocal}.

	\section{ FILTERED TDTMSV }

	\subsection{ Time-dependent Correlation Matrix of Filtered Output Modes}
	
	We consider another situation where a TMSV suffers decoherence due to continuous interaction with its thermal reservoir. The basic block diagram is shown in Fig. \ref{Block_Diagram_Thermal_decohr} where we see a TMSV field is thermally decorated after PDC before being filtered and detected. The interaction Hamiltonian between each mode of TMSV to its corresponding thermal reservoir is
	
	\begin{equation}
		H^{out}_{I,S} = \int_{-\infty}^{\infty} \eta^{out}_{I,S}(\omega) \left(a^{out}_{I,S} b^{out}_{I,S} (\omega)+h.c. \right) \mathrm{d}\omega
	\end{equation} 
	
	where $b^{out}_{I,S} (\omega)$ are the modes of the corresponding reservior, and $\eta^{out}_{I,S}(\omega)$ are the coupling strengths for the corresponding mode. For a wide band TMS state, the rate of thermalizations is determined by $\eta^{out}_{I,S}(\omega) = \sqrt{\kappa^{out}_{I,S}/\pi}$. 	
	The system in Fig. \ref{Block_Diagram_Thermal_decohr}, therefore estimates the correlation matrix elements mentioned in Eq. \eqref{correlation_matrix} as

	\begin{widetext}
		\begin{subequations}
			\begin{align}
				D_I &=  \left(\Im( {h}_K)^2+\Re( {h}_K)^2\right) \star \left( \Theta(t) ( {2n_I}+1) \left( 1- e^{-2 {\kappa_I} t}\right)+ ( \Theta(-t) + \Theta(t)e^{-2 {\kappa_I} t} ) \cosh (2 r)\right) \\
				D_S &= \left(\Im( {h}_L)^2+\Re( {h}_L)^2\right) \star \left(  \Theta(t) ( {2n_S}+1) \left( 1- e^{-2 {\kappa_S} t}\right)+ ( \Theta(-t) + \Theta(t) e^{-2 {\kappa_S} t} ) \cosh (2 r)\right) \\
				C_{11} &= -	C_{22}=\left(\Im( {h}_K) \Im( {h}_L)+ \Re( {h}_K) \Re( {h}_L) \right) \star ( \Theta(-t) + \Theta(t) e^{-t ({\kappa_I}+{\kappa_S})} ) \sinh (2r)\\
				C_{12} &=  0= C_{21}  
			\end{align}
		\end{subequations}
		
		which solves to
		
		\begin{align*}
			D_I &= ( {2n_I}+1) \left( I_I(0)- e^{-2 {\kappa_I} t} I_I(\kappa_I)\right)+ ( 1-I_I(0) + e^{-2 {\kappa_I} t} I_I(\kappa_I) ) \cosh (2 r) \\
			D_S &= ( {2n_S}+1) \left( I_S(0)- e^{-2 {\kappa_S} t}I_S(\kappa_S)\right)+ ( 1-I_S(0) + e^{-2 {\kappa_S} t} I_S(\kappa_S) ) \cosh (2 r) \\
			C_{11} &= -	C_{22}= \left( K_f - J_c(0) + e^{- ({\kappa_I}+{\kappa_S})t} J_c(({\kappa_I}+{\kappa_S})/2) \right) \sinh (2r)\\
			C_{12} &=  0= C_{21}  
		\end{align*}
	\end{widetext}
	
	Furthermore, using Eq. \eqref{entanglement_description} and Eq. \eqref{bell_func_deff}, we determine the entanglement and non-locality between the filtered modes.
	
	\subsection{Entanglement}
	
	The dissipation of entanglement of TMSV over time, and the impact of filter frequencies and linewidths on it, are plotted in Fig. \ref{LD_Ent_W_t} in a normalized time scale. We vary the filter frequencies, fixing their bandwidths in Fig. \ref{LD_Ent_W_t}(a)) for both the types of filters, and realized that the entanglement reduces uniformly with the mismatch of frequencies, throughout evolution. The phenomenon remains unchanged as it is seen during the evolution of TMSTDF in Fig. \ref{Ent_W_t}(a). 
	In  Fig. \ref{LD_Ent_W_t}(b), we changed the filter linewidths, fixing the central frequencies of both parties, and noticed that even though the mismatch of linewidths can eventually reduce entanglement, it does not remain consistent throughout the evolution. Increment of linewidth eventually reduces the effective rate of dissipation. 


	Furthermore, Fig. \ref{LD_Ent_R_n_k}(a) shows how the entanglement dissipates for different degrees of squeezing.  Unlike as observed in the case of TMSTDF, interestingly, we notice that the entanglement degrades faster for a higher degree of squeezing, for both types of filters. The phenomenon has already been noticed before in \cite{Two_mode_squeezed_vacuum_common_thermal_reservoir} where no filter was applied on TMSV suffering thermalization. This concludes that the application of a filter does not make any significant change in this situation. 	
	This ensures $r^{EN}_{max}$ to be reduced consistently throughout the evolution (Fig. \ref{LD_Ent_R_n_k}(b)). 
	Besides, as anticipated, the cutoff limits ($r^{EN}_{lcf}$ and $r^{EN}_{ucf}$) shrink the area of entanglement. Furthermore, likewise TMSTDF, we also notice that thermal dissipation of entanglement enhances for increasing thermal population (Fig. \ref{LD_Ent_R_n_k}(c)) and for reducing the rate of dissipation  (Fig. \ref{LD_Ent_R_n_k}(d)), in the normalized time scale.
	
	\subsection{ Non-Locality}

	Fig. \ref{LD_Bell_W_t} shows $|B|_{max}$ decreases when time passes by, and after reaching the minimum, $|B|_{max}$ increases to a value when it stabilizes. The dissipation of quantum non-locality of filtered TMSV, moreover follows the same profile of unfiltered TMSV, as it is observed in \cite{TMSV_nonlocal_Bell_Wigner}. Following entanglement, the
	non-locality changes uniformly with the mismatch of filter frequencies throughout evolution (Fig. \ref{LD_Bell_W_t}(a)), and non-uniformly with the mismatch of filter linewidth (Fig. \ref{LD_Bell_W_t}(b)).

	We also determined non-locality with the variation of the degree of squeezing in Fig. \ref{LD_Bell_R_n_k}(a), and noticed that the dissipation goes faster for higher $r$, for both the filters. The phenomenon has been witnessed and justified before for TMSV under thermalization in \cite{TMSV_nonlocal_Bell_Wigner}, where no filter was applied. It was proven that for higher $r$, the superposition of two coherent states can easily be destroyed.
	The phenomenon remains visilant even after applying filter. 
	This leads to observing the input squeezing for maximal non-locality ($r^B_{max}$) to go down with time in Fig. \ref{LD_Bell_R_n_k}(b), and both the cutoff limits ($r^B_{lcf}$ and $r^B_{ucf}$) show the region of non-locality to shrink with time. Finally, Fig  \ref{LD_Bell_R_n_k}(c,d) shows, in a normalized dimensionless time scale, the dissipation of the Bell function increases with the increment of thermal population and the reduction of coupling with the environment, irrespective of the type of filter.

	\section{CONCLUSION}

	We investigated the effects of filters on the dynamics of entanglement and the non-locality of thermally decorated two-mode squeezed (TMS) states. Both types of thermal decoherence -- TMSTDF and TDTMSV - were considered before applying the filter and detector. The system was analyzed using two distinct types of filters -- step and exponential. Our findings reveal that entanglement and Bell measurement remain maximized when identical filters are used, with reductions in entanglement and non-locality displaying symmetric behavior as the central frequencies of the filters mismatch, remaining consistent with previous observations in steady states in \cite{mypaper_TMSV_filter}. However, when there is a mismatch in the linewidths of the filters, the reduction no longer remains symmetric. 

	It is anticipated that, after an initial increase, entanglement and Bell measurements are expected to decline as the input squeezing degree increases, forming a bell-shaped pattern \cite{mypaper_TMSV_filter}. However, in the case of TMSTDF, this behavior does not remain consistent throughout the evolution. Interestingly, the entanglement becomes more robust for higher $r$, during the period of thermal dissipation, even though initial and final entanglement remains weaker. For higher $r$ the dissipation starts slower and gets faster over time. However, the non-locality moreover decays at the same rate irrespective of initial squeezing. 
	In the case of TDTMSV, dissipation of both entanglement and non-locality intensifies for higher values of $r$ \cite{Two_mode_squeezed_vacuum_common_thermal_reservoir, TMSV_nonlocal_Bell_Wigner}. The trend persists even after applying the filter. This therefore makes the region of entanglement and non-locality to shrink down which is indicated by their cutoff limits. Furthermore, in all cases, the rate of dissipation increases with a higher thermal population but decreases as coupling strength increases on a normalized, dimensionless time scale.

	Notably, in mixed states, while entanglement is necessary for non-locality, it is not always sufficient. The rapid degradation of entanglement and non-locality with higher degrees of input squeezing raises concerns about the practicality of generating highly squeezed TMS states for use in quantum optical experiments, particularly those intended for quantum communication or gravitational wave metrology \cite{entanglement_quantum_cryptography, quantum_computation, quantum_teleportation, study4roadmap}.

	\begin{acknowledgments}
		SA would like to thank Philippe Djorwe for his fruitful suggestions. The work has been supported by the European Union, MSCA GA no 101065991 (SingletSQL).
		
	\end{acknowledgments}

	\appendix
	
	\section{ Determination of Correlation Matrix Elements  }\label{Correlation_Matrix_Elements}

	\subsection{Filter Type - I}

	The elements of the matrix $V (r;t)$ can be determined in the time domain by solving the following integrals.
	
	\begin{widetext}
		\begin{align*}
			& \int_{-\infty}^{t} \left[\Re({h}_K (t-s)) 	\Im({h}_L(t-s)) - \Im({h}_K (t-s) ) \Re({h}_L (t-s)) \right] n \Theta(t) \exp (-2 \kappa  s) \,\, ds \\
			& = \frac{1}{\sqrt{\tau _I \tau _S}} \int_{t-\tau}^t \sin \left[(t-s) \left( \Omega _K-\Omega _L \right)\right]  n \,\,\exp (-2 \kappa  s) \, ds \\
			& = n  e^{-2 \kappa  t} J_s(\kappa) 
		\end{align*}

		\begin{align*}
			&  \int_{-\infty }^{t} \left[\Re({h}_K (t-s)) 	\Re({h}_L(t-s)) + \Im({h}_K (t-s) ) \Im({h}_L (t-s)) \right]  n \Theta(t) \, \exp (-2 \kappa  s)\,\, ds\\
			& = \frac{1}{\sqrt{\tau _I \tau _S}} \int_{t-\tau}^t \cos \left[(t-s) \left(\Omega _K-\Omega _L \right)\right] n\,\,\exp (-2 \kappa  s)  \,\, ds \\
			& = n  e^{-2 \kappa  t} J_c(\kappa) 
		\end{align*}
		
	\end{widetext}
	
	where $\tau = \min [t,\tau_{I}, \tau_{S}]$.
	When both the filters become identical, i.e. $\Omega_K = \Omega_L$ and $\tau_I = \tau_S$; $J_c$ determines the integrals $J_c = I_I$  or $J_c = I_S$. Also, one can check that $K_f = J_c(0)$ when $t > \max[\tau_{I}, \tau_{S}] $.

	\subsection{Filter Type - II}
	
	The integrals can be determined for the filter of type-II as
	
	\begin{widetext}
		\begin{align*}
			& \int_{-\infty}^{t} \left( \Re({h}_k) 	\Im({h}_s)(t-s) - \Im({h}_k)\Re({h}_s)(t-s) \right) \left(n \Theta(t) (e^{ (-2 \kappa  s)}) \right) ds \\
			& \int_{-\infty}^t \frac{e^{-(t-s)/\tau_{I} } e^{-(t-s)/\tau_{S} } }{\sqrt{\tau_{I} \tau_{S}/4}}  \sin \left[(t-s) (\Omega _K-\Omega _L)\right] \left(n \Theta(t) (e^{ (-2 \kappa  s)}) \right)   \, ds \\
			& = n  e^{-2 \kappa  t} J_s(\kappa)  \\
		\end{align*}

		\begin{align*}
			& \int_{-\infty}^{t} \left( \Re({h}_k) 	\Re({h}_s)(t-s)+ \Im({h}_k)\Im({h}_s)(t-s) \right) \left(n \Theta(t) (e^{ (-2 \kappa  s)}) \right) ds \\
			& \int_{-\infty}^t \frac{e^{-(t-s)/\tau_{I} } e^{-(t-s)/\tau_{S} } }{\sqrt{\tau_{I} \tau_{S}/4}}  \cos \left[(t-s) (\Omega _K-\Omega _L)\right] \left(n \Theta(t) (e^{ (-2 \kappa  s)}) \right)   \, ds \\
			& = n e^{-2 \kappa  t} J_c(\kappa)  \\
		\end{align*}

	\end{widetext}
	
	Similar to filter-I, when both the filters become identical, i.e. $\Omega_K = \Omega_L$ and $\tau_I = \tau_S$; $J_c$ determines the integrals $J_c = I_I$  or $J_c = I_S$. In this case $K_f = J_c(0)$ when $t \to \infty$.
	
	--------------------------------

	\nocite{*}

	\bibliography{apssamp}
	\bibliographystyle{apsrev4-2}

\end{document}